\documentclass[aps,prd,amsmath,amssymb]{revtex4}

\usepackage{graphicx}
\usepackage{dcolumn}
\usepackage{bm}

\newcommand{\be}{\begin{equation}}
\newcommand{\ee}{\end{equation}}
\newcommand{\bea}{\begin{eqnarray}}
\newcommand{\eea}{\end{eqnarray}}
\newcommand{\hf}{\frac{1}{2}}
\newcommand{\nn}{\nonumber\\}
\def\journal#1#2#3#4{{#1}{\bf#2},#3(#4)}
\def\eq#1{(\ref{#1})}
\def\la{\langle}
\def\ra{\rangle}

\def\tr{{\mathrm Tr}}
\def\ord#1{{\cal O}\left(#1\right)}

\def\part#1#2{\frac{\partial#1}{\partial#2}}

\def\gb{\bar g}

\begin{document}
\title{Casimir effect: running Newton constant or cosmological term}
\author{Janos Polonyi$^{a,b}$}\email{polonyi@fresnel.u-strasbg.fr}
\author{Enik\H o Reg\H os$^{b,c}$}\email{eniko@ast.cam.ac.uk}
\affiliation{$^a$Theoretical Physics Laboratory, CNRS and
Louis Pasteur University, Strasbourg, France}
\homepage{http://lpt1.u-strasbg.fr}
\affiliation{$^b$Department of Atomic Physics,
L. E\"otv\"os University, Budapest, Hungary}
\affiliation{$^c$Theoretical Physics Group, Hungarian Academy of Sciences,
Budapest}

\date{\today}

\begin{abstract}
We argue that the instability of Euclidean Einstein gravity
is an indication that the vacuum is non perturbative and
contains a condensate of the metric tensor in a manner reminiscent of
Yang-Mills theories. As a simple step
toward the characterization of such a vacuum the value of the
one-loop effective action is computed for Euclidean de Sitter
spaces as a function of the curvature when the unstable conformal
modes are held fixed. Two phases are found, one where the curvature
is large and gravitons should be confined and another one which
appears to be weakly coupled and tends to be flat.
The induced cosmological constant is positive or negative in
the strongly or weakly curved phase, respectively.
The relevance of the Casimir effect in understanding the UV sensitivity
of gravity is pointed out.
\end{abstract}
\maketitle

\section{Introduction}
Quantum Gravity possesses a number of fundamental questions which are
unsolved so far. Its apparent conflict with Quantum Mechanics, namely
canonical commutation relations imposed on equal time
hypersurface require the knowledge of the metric before the
operator equations defining it is completed, questions the
need of the quantum treatment of gravitation altogether \cite{adler}.
Let us assume that the metric corresponds either to an elementary
or to a composite field and the usual quantum field theoretical
path integral quantization applies. The next problem
arises from the UV divergences. At this point one either
seeks some superstring formalism which might be interpreted
as a physical regulator or relies on a formal regulator. We choose the
latter being a simpler possibility and do not attempt to remove
the cutoff, i.e. we are not interested in the problem of
physics at arbitrary high energies.
Instead, our goal in this work is to make a small step toward
the understanding of the nature of the radiative corrections
up to a high but finite energy for a particular background geometry,
those of the de Sitter space.

But there are still problems even after such a drastic reduction of
the scope of the investigation.
Regulators which preserve the local symmetries of the theory
and allows analytical computations are available for the
Euclidean section of the space-time only.
When Euclidean Quantum Gravity is considered then some
or all of the conformal modes become unstable \cite{instab},
rendering the perturbative or saddle point approximation problematical.
The usual approach to this complication is to stabilize
Euclidean Quantum Gravity on the tree level by an appropriate
modification of the Wick rotation \cite{wickqg} or on the quantum level
by altering the rules of quantization \cite{quantst} or the
functional measure for the path integral \cite{geom}.

Similar instability is well known in Yang-Mills models.
In fact, Euclidean Yang-Mills theory considered in the presence
of a homogeneous magnetic field has unstable modes \cite{savvidy}.
Since the Euclidean Yang-Mills action
is bounded from below the Euclidean vacuum is stable
though the detailed mechanism of this stabilization is still unknown.
When the time extent of the Euclidean space can be sufficiently
large then the large amplitude modes resulting from the instability
can be regarded as an indication of a condensate of the true,
physical vacuum of the theory with real time.
Can this lesson be relevant for Quantum Gravity and pointing toward
the non triviality of its vacuum?

One can imagine three dynamical, consistent stabilization
mechanisms for Euclidean Quantum Gravity. (i) It might happen that
the true gravitational action contain higher order derivative
terms which renders the classical action bounded from below.
If these new terms do not involve new dimensional coupling constant
then the stable vacuum will be dominated by fluctuations with scale
close to the cutoff because the instability of the
Euclidean Einstein action is in the UV regime. (ii) Another possibility is that the classical action is unbounded but
the theory is stabilized by quantum fluctuations. This happens
if the classical instability tries to squeeze the system into
a small region of the configuration space and the kinetic energy
generated by the uncertainty principle becomes strong.
Certain numerical simulations seem to support this scenario
for quantum gravity regularized by the Regge-calculus \cite{lattice}.
(iii) Yet another possibility is that pure gravity is indeed unstable
but the stabilization is achieved by the coupling
of gravity to matter \cite{axanom,conform}.

But one should bear in mind that either possibility is realized the
vacuum is non perturbative and contain a condensate. If the vacuum
is stabilized on the $n$-th order of the loop expansion then
it corresponds to the minimum of the effective action computed
up to $\ord{G^n_B}$. Here $G_B$ denotes the bare Newton constant which
plays the role of $\hbar$ in the absence of matter and counts the order
of the loop expansion. The determination of the minimum of this effective
action requires the knowledge of all 1PI vertex functions
in the given order of the loop expansion. In the case of $n=0$,
c.f. case (i), the solution of the classical equation of motion
can be regarded as given by summing over infinitely many
tree graphs. When $n>0$, c.f. cases (ii) and (iii), then an infinite
sum of graphs up to $n$-loops gives the vacuum configuration.

We do not attempt to identify the true stabilization mechanism
in this paper. Instead, we shall be contended with a small step
along the line of reasoning outlined above and present a study
of the dynamical environment in which the stabilization takes place by
computing the effective potential for a homogeneous
space-time curvature with vanishing unstable modes and
by reading off from it the running of the Newton
constant with the length scale of such a background geometry.

Before we continue the discussion of the computation of the effective
potential let us bring another pertinent element,
the Casimir effect, into the discussion. The vacuum energy
of quantum field theories in flat space-time depends
on a rather complicated manner on the choice of the
space-time region $\cal R$ populated by quantum fields
\cite{casres}. It is well known that the metric
tensor controls all aspects of the geometry, among them
the spatial extent of the system. In a space-time of
homogeneous, positive curvature the free geodesics are
deflected in such a manner that they stay in a
region with linear size in the order of
\be\label{circ}
L\approx\frac1{\sqrt{R}}.
\ee
Such double role of the curvature, being an external field
and an IR cutoff in the same time, provides a new, physical insight
into the inherent sensitivity of quantum gravity on the UV
sector. The argument is the following.
The partition function computed on a homogeneous or at least slowly
varying background field in a fixed space-time region $\cal R$
develops UV finite dependence on the background field after
having introduced the usual background field independent counterterms.
One would expect the same conclusion when the partition function
is considered on a background curvature. But this is not what happens
because of the Casimir effect. The reason is that the Casimir
effect is always UV divergent. To understand this property
let us consider first the number of modes
\be\label{modusok}
{\cal N}\approx(\Lambda L)^d
\ee
in dimension $d$ where $\Lambda$ denotes the UV cutoff
of dimension energy and $L$ stands for the linear size of
the system. The expression \eq{modusok} diverges when the UV and IR
cutoffs are removed because the length scale
of a mode can be infinitely large or infinitely small.
The result is that the UV and the IR dependences are
mixed in $\cal N$. In a similar manner the Casimir effect
mixes the UV and the IR sectors, too. Therefore the
dependence of the partition function on the curvature
is always UV divergent. It is worthwhile noting that this point is
nothing but the throughly studied issue of the coupling of
the zero point fluctuations of the vacuum to the gravitation.

Our main goal in this work is to determine the renormalization of the
effective potential for the curvature and some effective coupling
constants in one-loop order.
The effective potential, $\gamma(R)$, is defined by the
value of the effective action for homogeneous de Sitter spaces of
curvature $R>0$ and will be computed in $\ord{G_B}$
by restricting the unstable modes to zero, except the homogeneous
conformal mode which is kept constant. We do not attempt to remove the
UV cutoff and confine our investigation on the finite
energy dynamics by keeping $\Lambda$ fixed and ignoring the
formal issue of renormalizability. We find a phase transition
when the 'best' de Sitter space is selected by minimizing $\gamma(R)$.
For low enough value of the cutoff, $\Lambda<<\Lambda_{cr}$ the
energetically preferred value of the curvature is $\ord{\Lambda^4G}$.
The theory is weakly coupled at such or larger values of the curvature
but the modes with length scale $\ell>2/\Lambda^2\sqrt{G}$ are non
perturbative. The other phase, realized by large cutoff
$\Lambda>>\Lambda_{cr}$, prefers small curvature and is perturbative.
This phase structure agrees qualitatively with the one found in
lattice simulations \cite{lattice} where a flat and a strongly
curved phase were found for large or small values of the
bare Newton constant, respectively.
One should note in this respect that although the UV cutoff represents an
upper limit for the curvature its relation to the Newton constant
can be arbitrary, i.e. quantum gravity is a well defined and physically
nontrivial theory for either $G\Lambda^2<1$ or $G\Lambda^2>1$.

The Casimir contribution to the effective action is proportional to
the space-time volume which raises the possibility of generating a
cosmological constant even in the absence of matter. In fact, an
induced cosmological constant is found whose sign distinguishes the two
phases mentioned above.

The UV cutoff, $\Lambda$, appears as an unavoidable parameter in the
main results of this paper, in the renormalization of the effective
action and the phase structure. Such a state of affairs is certainly
erroneous in the case of a renormalizable theory. But one should remember
that whatever is the ultimate theory of quantum gravity the quantum
effects are supposed be governed by the Einstein-Hilbert action within
the energy and distance range explored up to today.
The predictions of this theory, being nonrenormalizable, naturally do
depend in an essential manner on the choice of the UV cutoff.
But this dependence is restricted to scales
comparable to the cutoff. At energy scales far below the
cutoff quantum gravity is a useful effective theory because
the nonrenormalizable, UV irrelevant terms of the action
play no role here \cite{effth}. It is worthwhile noting that
this energy regime is determined by the cutoff only, the
other dimensional parameter of the theory discussed in this
work, the Newton constant, may take arbitrary values with
respect to the UV cutoff in this effective gravity.

There have already been investigation of effective coupling constants
carried out before. The one-loop correction to the Newtonian
gravitational potential on flat space-time has been computed
in the IR domain in Refs. \cite{effnc} and a weak, $\ord{R^{-2}}$,
UV finite effect was found. The result is expected to be rather
different when higher loop contributions are retained because
those graphs contain UV divergences. The self interaction
of gravitons, considered in our computation, is UV divergent
already at the one-loop level and the comparison of the
running Newton constant presented in these works with those
arising from the computation of the gravitational forces between
matter particles is not obvious. In fact, the difference of the
dependence on the scale of a space-time variable or of a field amplitude
is usually a manifestation of a nontrivial wave function renormalization
constant \cite{int}. The non triviality of the wave function renormalization 
constant can be expected on the simple dimensional ground that the
metric tensor is a dimensionless field as opposed to the usual one.
The conventional renormalization group has already been applied 
to determine the running of the coupling constants of Grand Unified 
Theories on curved geometry and a number of scaling laws were
identified \cite{convrg}. But one should bear in mind that the multiplicative
renormalization which serves as the basis for the conventional
renormalization group equation is valid for renormalizable models
only. The renormalization group flow was computed for the Newton and the 
cosmological constants in Ref. \cite{reuter} for de Sitter spaces. Two kinds of fixed
points were found, a Gaussian and gauge-dependent non-Gaussian.
The flow was extended to arbitrary dimension and value of the
gauge fixing parameter in Ref. \cite{litim}.
But it is difficult to compare thess result with ours, partly because this
computation involves a formal one-loop integral over the unstable
modes, partly because of the presence of the cosmological constant.
The gauge dependence of the renormalization group flow was investigated
in Ref. \cite{gaugdep}. Though our motivation is different,
our work is closest in technics to Ref. \cite{venez} where
the dynamical suppression of the cosmological constant was
studied in de Sitter spaces.

The organization of the paper is the following. The effective
potential and the technical details such as the regularization,
gauge fixing and the eigenfunctions of the Laplace operator
in de Sitter spaces are introduced in Section II. The results
of the numerical computation are presented in Section III.
The problem of the stability of the conformal modes is discussed in
Section IV. Finally, Section V is for the summary and conclusions. A very brief
summary of the simplest appearance of the Casimir effect is
presented in the Appendix.

\section{Effective potential for the curvature}
The computation of the value of the one-loop effective potential for
de Sitter spaces is presented in this section.

\subsection{Formal expressions}
The generator functional for connected Green functions in
the Euclidean section is defined as
\be\label{bfpi}
e^{W[j]}=\int D[g]e^{-S[g]+\int dxj^{\mu\nu}g_{\mu\nu}},
\ee
where the Einstein-Hilbert action is given in terms of the metric tensor $g$,
\be
S[g]=-\kappa^2_B\int dx\sqrt{g}(R-2\lambda_B),
\ee
the multiplicative constant in front of the action
is $\kappa^2_B=1/16\pi G_B$ and $\lambda_B$ denotes the cosmological constant
which will be left vanishing on the tree-level for simplicity. 
We shall use units $c=\hbar=1$.
The effective potential for the metric, the Legendre transform of $W[j]$,
\be\label{effaa}
\Gamma[g]=-W[j]+\int dxj^{\mu\nu}g_{\mu\nu}
\ee
is introduced by the definition
\be\label{geffaa}
g=\frac{\delta W[j]}{\delta j}.
\ee
Finally, an effective potential for the homogeneous curvature
will be identified as the value of the effective action at a
homogeneous curvature,
\be
\gamma(R)=\Gamma[g^{(R)}]
\ee
where $g^{(R)}$ denotes the metric tensor of the de Sitter space
of curvature $R$. Notice that $\gamma(R)$ is not the conventional
effective potential corresponding to the composite operator $R$,
their difference arises from the
fluctuations of the metric when the curvature is held fixed.
We shall use the simpler construction, $\gamma(R)$, because
our goal is the find the geometry which minimizes the effective action
under the assumption of the homogeneity of the space-time.

The conformal modes are instable in the Euclidean theory \cite{instab}.
In fact, the transformations
\be\label{resc}
g_{\mu\nu}\to\Omega^2g_{\mu\nu},~~~
R\to\frac{R}{\Omega^2}-6\frac{\Box\Omega}{\Omega^3},
\ee
induce the change
\be\label{actcon}
S[g]\to S[g]=-\kappa^2\int dx\sqrt{g}
\left(6D_\mu\Omega D_\nu\Omega g^{\mu\nu}+\Omega^2 R
-2\lambda_B\Omega^4\right)
\ee
showing that the action is unbounded from below for
inhomogeneous $\Omega$. In order to isolate the instability the generating functional
will be computed in two consecutive steps, first we integrate over the stable modes
and the conformal modes will be dealt with in the second step.

The one-loop approximation yields
\be
W[j]=-S[\gb]+\int dx\sqrt{g}j^{\mu\nu}\gb_{\mu\nu}
+\hf\tr\ln\frac{\delta^2S[\gb]}{\delta g\delta g}
\ee
where $\gb$ is the saddle point satisfying the classical
equation of motion
\be
\frac{\delta S[\gb]}{\delta g}=j
\ee
and $g=\gb+h$ with $h=\ord{\kappa^{-2}_B}$. As a result we find
\be
\Gamma[g]=S[g]+\hf\tr\ln\frac{\delta^2S[g]}{\delta g\delta g},
\ee
in particular for $\gb=g^{(R)}$,
\be\label{efpotr}
\gamma(R)=S[g^{(R)}]+\hf\tr\ln\frac{\delta^2S[g^{(R)}]}{\delta g\delta g},
\ee
up to terms $\ord{\kappa^{-2}_B}$.

One has to settle two problems in order the manipulation outlined above
be well defined. One needs gauge invariant UV regulator and the unstable modes 
must be kept under control in the functional integration.

\subsection{Gauge fixing and regularization}
In order to saturate $W[j]$ by a Gaussian integral one has
to fix the gauge. We shall use harmonic gauge by adding
\be
S_{gf}[g^{(R)},h]=\hf\int dx\sqrt{g^{(R)}}g^{(R)\mu\nu}
f_\mu[g^{(R)},h]f_\nu[g^{(R)},h]
\ee
to the action where
\bea
f_\mu[g^{(R)},h]&=&\sqrt{2}\kappa_Bf_\mu^{\nu\rho}[g^{(R)}]h_{\nu\rho},\nn
f_\mu^{\alpha\beta}[g^{(R)}]&=&\hf\delta_\mu^\alpha
g^{(R)\beta\gamma}D_\gamma+\hf\delta_\mu^\beta g^{(R)\alpha\gamma}D_\gamma
-\hf g^{(R)\alpha\beta}D_\mu,
\eea
and $D$ stands for the covariant derivative of the metric $g^{(R)}$.
The standard Fadeev-Popov quantization rules lead to the
generator functional
\be
e^{W[j]}=\int D[h]D[c]D[\bar c]
e^{-S[g^{(R)}+h]-S_{gf}[g^{(R)},h]+S_{gh}[g^{(R)},h,c,\bar c]
+\int dx\sqrt{g}j^{\mu\nu}(g^{(R)}_{\mu\nu}+h_{\mu\nu})}
\ee
where $c$ and $\bar c$ denote the anticommuting ghost fields and
\be
S_{gh}[g^{(R)},h,c,\bar c]=-\sqrt{2}\int dx\sqrt{g^{(R)}}
\bar c^\mu{\cal M}_{\mu\nu}[g^{(R)},h]c^\nu
\ee
with
\be
{\cal M}_{\alpha\beta}=\sqrt{2}\kappa_Bf_\alpha^{\mu\nu}[g^{(R)}]
\left(\part{g_{\mu\nu}}{x^\beta}-\part{g_{\beta\nu}}{x^\mu}
-\part{g_{\mu\beta}}{x^\nu}\right).
\ee
Notice that $S_{gh}$ is independent of the gauge invariant cosmological 
constant because it is determined by the properties of the gauge orbits
in the configuration space. In the same time the dynamics of
the metric tensor naturally contains $\lambda_B$. This is reminiscent
of the mass generation by spontaneous symmetry breaking where the
mass term for the gauge field is gauge invariant, too.

The UV regulator will be constructed by
a gauge invariant generalization of the momentum cutoff.
Since the eigenvalues of the covariant derivative are gauge invariant
the theory can be regulated in a non perturbative manner
by using the bare action
\be\label{regul}
S_B[g]=S[g]+\int dx\sqrt{g}g_{\mu\nu}f_\Lambda(-D^2)g^{\mu\nu},
\ee
where $f_\Lambda(z)\approx0$ and $\infty$ for $z<\Lambda^2$ and
$z>\Lambda^2$, respectively.
Smooth cutoff is realized by either the proper time method \cite{schw}
or by other, simpler continuous function $f_\Lambda$.
The simplest possibility, followed in this work is the sharp cutoff,
\be\label{sharp}
f_\Lambda(z)=\begin{cases}0&z<\Lambda^2,\cr\infty&z>\Lambda^2.\end{cases}
\ee

The expression for the regulated one-loop effective potential for the
curvature \cite{reuter} can be obtained by following the steps leading
from Eq. \eq{bfpi} to Eq. \eq{efpotr} with the regulated and gauge fixed
path integral,
\bea\label{effpot}
\gamma(R)&=&S_B[g^{(R)}]+\gamma^{(1)}(R),\nn
\gamma^{(1)}(R)&=&\frac{\kappa^2_B}{2}\tr\ln{\cal K}-\tr\ln{\cal M},
\eea
with
\bea\label{qe}
{\cal K}_{(\mu\nu),(\rho\sigma)}
&=&\kappa^2_B\sqrt{g^{(R)}}\left(-Z_{(\mu\nu),(\rho\sigma)}D^2
+U_{(\mu\nu),(\rho\sigma)}\right),\nn
{\cal M}_{\mu\nu}&=&-\sqrt{2}\kappa^2_B\sqrt{g^{(R)}}
\left(g^{(R)}_{\mu\nu}D^2+R_{\mu\nu}\right),
\eea
where
\bea
Z_{(\mu\nu),(\rho\sigma)}&=&\frac14\left(g^{(R)}_{\mu\rho}g^{(R)}_{\nu\sigma}
+g^{(R)}_{\mu\sigma}g^{(R)}_{\nu\rho}-g^{(R)}_{\mu\nu}g^{(R)}_{\rho\sigma}\right),\nn
U_{(\mu\nu),(\rho\sigma)}&=&Z_{(\mu\nu),(\rho\sigma)}(R-2\lambda_B)
+\hf\left(g^{(R)}_{\mu\nu}R_{\sigma\rho}+R_{\mu\nu}g^{(R)}_{\sigma\rho}\right)
-\frac14\left(g^{(R)}_{\mu\rho}R_{\nu\sigma}+g^{(R)}_{\mu\sigma}R_{\nu\rho}
+R_{\mu\rho}g^{(R)}_{\nu\sigma}+R_{\mu\sigma}g^{(R)}_{\nu\rho}\right)\nn
&&-\hf\left(R_{\mu\rho}R_{\nu\sigma}+R_{\mu\sigma}R_{\nu\rho}\right),
\eea
and $R_{\mu\nu}$ is the Ricci tensor of the de Sitter space.

The evaluation of the functional traces on the right hand side
will be carried out in the two steps. First we project the fields
into subspaces with well defined spin in order to simplify the
computation of the spectrum of $\cal K$ and $\cal M$. In the
second step we construct the spherical harmonics for each
spin sector and obtain the functional traces as regulated, finite
sums.

\subsection{Spin projection}
The computation of the eigenvalues of the quadratic forms
of Eq. \eq{qe} is facilitated by separating
the different spin components of the fields $h_{\mu\nu}$,
$c_\mu$ and $\bar c_\mu$. In the case of the metric the reduction
produces traceless transverse $(TT)$ and longitudinal
$(LT),(LL)$ fields and the trace $(Tr)$ \cite{ttr},
$10=2\times1(LL+Tr)+3(LT)+5(TT)$, in particular
\be\label{reduct}
h_{\mu\nu}=h^{TT}_{\mu\nu}+h^{LT}_{\mu\nu}+h^{LL}_{\mu\nu}+h^{Tr}_{\mu\nu}
\ee
with $h^{Tr}_{\mu\nu}=g^{(R)}_{\mu\nu}\phi$ where
$\phi=g^{(R)}_{\mu\nu}h^{\mu\nu}$
together with $h^{LT}_{\mu\nu}=D_\mu\xi'_\nu+D_\nu\xi'_\mu$
and $h^{LL}_{\mu\nu}=D_\mu D_\nu\sigma'-\frac14g^{(R)}_{\mu\nu}D^2\sigma'$.
These fields satisfy the auxiliary conditions $D_\mu\xi^\mu=0$,
$g^{(R)\mu\nu}h^{TT}_{\mu\nu}=0$ and $D^\mu h^{TT}_{\mu\nu}=0$
and are orthogonal with respect to the scalar product \cite{geom,ttr}
\bea\label{sclor}
\la h_1|h_2\ra&=&\int dx\sqrt{g}h_{1;\mu\nu}h^{'\mu\nu}_2\nn
&=&\int dx\sqrt{g}\biggl[h^{TT}_{1;\mu\nu}h^{TT'\mu\nu}_2
-2\xi'_{1;\mu}(g^{(R)\mu\nu}D^2+\bar R^{\mu\nu})\xi'_{2;\nu}
-2\xi'_{1;\mu}\bar R^{\mu\nu}D_\nu\sigma'_2
-2\xi'_{1;\mu}\bar R^{\mu\nu}D_\nu\sigma'_2\nn
&&+\sigma'_1\left(\frac34(D^2)^2+D_\mu\bar R^{\mu\nu}D_\nu\right)\sigma'_2
+\frac14\phi_1\phi_2\biggr]
\eea
where $g^{(R)\mu\rho}g^{(R)\nu\sigma}h_{\rho\sigma}=h^{\mu\nu}$.
In order remove a trivial divergent factor from the
path integral the zero modes $D_\mu\xi_\nu+D_\nu\xi_\mu=0$
and $D_\mu D_\nu\sigma-\frac14D^2\sigma=0$ should
be removed from the domain of integration in \eq{bfpi}.
The comparison of the rescaling \eq{resc}
and the decomposition \eq{reduct} reveals
\be\label{conff}
\Omega^2=1+\phi.
\ee

For maximally symmetric spaces $\bar R^{\mu\nu}=Cg^{(R)\mu\nu}$ and the
$\xi-\sigma$ mixing is absent in the scalar product,
\be
\la h_1|h_2\ra=\int dx\sqrt{g}\biggl[h^{TT}_{1;\mu\nu}h^{TT'\mu\nu}_2
-2\xi'^\mu_1(D^2+C)\xi'_{2;\mu}
+\sigma_1'D^2\left(\frac34D^2+C\right)\sigma'_2
+\frac14\phi_1\phi_2\biggr]
\ee
A further linear transformation \cite{reuter},
\be
\tilde\xi=\sqrt{-D^2-C}\tilde\xi',~~~
\tilde\sigma=\sqrt{(D^2)^2+\frac43CD^2}\tilde\sigma'
\ee
simplifies the scalar product to
\be\label{scpro}
\la h_1|h_2\ra=\int dx\sqrt{g}\biggl[h^{TT}_{1;\mu\nu}h^{TT'\mu\nu}_2
+2\xi^\mu_1\xi_{2;\mu}+\frac34\sigma_1\sigma_2+\frac14\phi_1\phi_2\biggr].
\ee
Similar reduction of the ghost fields gives $4=1+3$,
$c_\mu=c^T_\mu+D_\mu(-D^2)^{-\hf}\rho$ and
$\bar c_\mu=\bar c^T_\mu+D_\mu(-D^2)^{-\hf}\bar\rho$,
where $D_\mu c^{T\mu}=0$.

The comparison of the first equation in \eq{sclor}
and \eq{scpro} reveals that the Jacobians
corresponding to the change of integration variables
$h\to\Phi$, where $\Phi\in\{h^{TT},\xi,\sigma,\phi,c^T,\bar c^T,\rho,\bar\rho\}$
is field and curvature independent.
The quadratic part of the actions, given by Eq. \eq{qe} reads as
\bea\label{quadr}
S^{(2)}[g^{(R)}+h]+S_{gf}[g^{(R)}+h]&=&\kappa^2_B\int dx\sqrt{\bar g}\Biggl[
\hf h^{TT}_{\mu\nu}\left(-D^2+\frac{2R}{3}-2\lambda_B\right)h^{TT\mu\nu}\nn
&&+\xi_\mu\left(-D^2+\frac{R}{4}-2\lambda_B\right)\xi^\mu
+\frac38\sigma(-D^2-2\lambda_B)\sigma-\frac16\phi(-D^2-2\lambda_B)\phi\Biggr]
\eea
and
\be
S_{gh}[g^{(R)},h=0,c^T,\bar c^T,\rho,\bar\rho]
=-\sqrt{2}\kappa^2_B\int dx\sqrt{g}\biggl[
\bar c^T_\mu\left(D^2+\frac{R}{4}\right)c^{T\mu}
+\bar\rho\left(D^2+\frac{R}{4}\right)\rho\biggr]
\ee
in terms of the new variables. Notice that all $\phi$ modes are
unstable when $\lambda_B=0$ and will be dealt with later.

\subsection{Spherical harmonics}
The last ingredient needed for the one-loop result is the
spectrum of the Laplace operator $D^2$, the solution of the
eigenvalue condition
\be\label{sphha}
D^2t^{(s)}_{\ell m}=-R\lambda_\ell^{(s)}t^{(s)}_{\ell m},
\ee
for spherical functions of spin $s$ where $\ell=s,s+1,s+2,\cdots$
and the quantum number $m=1,\cdots,D^{(s)}_\ell$ labels the
degenerate eigenvectors.
The spherical functions can be constructed either from
homogeneous polynomials \cite{poly} or in terms of
associated Legendre polynomials \cite{leg} with the spectrum
\bea
\lambda_\ell^{(s)}&=&\frac{\ell(\ell+3)-s}{12},\nn
D_\ell^{(s)}&=&\begin{cases}
\frac{(2\ell+3)(\ell+1)(\ell+2)}{6}&s=0(LL,Tr),\cr
\frac{(2\ell+3)\ell(\ell+3)}{2}&s=1(LT),\cr
\frac{5(\ell+4)(\ell-1)(2\ell+3)}{6}&s=2(TT).\end{cases}
\eea

The fields are expressed by means of the spherical harmonics as
\be
\Phi=\sum_{\ell=\ell^\Phi_{min}}^{\ell^\Phi_{max}}\sum_{m=1}^{D^\Phi_\ell}
c^\Phi_{\ell m}t^\Phi_{\ell m}.
\ee
The spherical harmonics $t^\Phi_{\ell m}$ are normalized by the scalar product
\be
\la t_1|t_2\ra=\int dx\sqrt{g}t^*_1t_2,
\ee
the lower limit for the summation over $\ell$ is $\ell^\Phi_{min}=2$
except for $\Phi=\phi$ when $\ell^\phi_{min}=0$.
The upper limit is determined by the sharp cutoff, Eq. \eq{sharp}, therefore
the integral measure for $c_n$ is finally defined as
\be
D[\Phi]\equiv D[c^\Phi]=\prod_{\ell=\ell^\Phi_{min}}^{\ell^\Phi_{max}}
\prod_{m=1}^{D^\Phi_\ell}\Lambda^2d\tilde c^\Phi_{\ell m}.
\ee
The final expression for the effective potential \eq{effpot} is
\bea\label{sumsph}
\gamma^{(1)}(R)&=&\hf\sum_{\ell=2}^\infty D_\ell^{(2)}\ln\left[
\frac{\kappa^2_BR}{\Lambda^4}\left(\lambda^{(2)}_\ell+\frac{2}{3}
-\frac{2\lambda_B}{R}\right)\right]+\hf\sum_{\ell=2}^\infty D_\ell^{(1)}\ln\left[
\frac{\kappa^2_BR}{\Lambda^4}\left(2\lambda^{(1)}_\ell+\hf-\frac{4\lambda_B}{R}\right)\right]\\
&&\hskip-4pt+\hf\sum_{\ell=2}^\infty D_\ell^{(0)}\ln\left[\frac{3\kappa^2_BR}{4\Lambda^4}
\left(\lambda^{(0)}_\ell-\frac{2\lambda_B}{R}\right)\right]
-\sum_{\ell=1}^\infty D_\ell^{(1)}\ln\left[\frac{\kappa^2_BR}{\Lambda^4}
\left(\lambda^{(1)}_\ell+\frac{1}{4}\right)\right]-\sum_{\ell=0}^\infty D_\ell^{(0)}
\ln\left[\frac{\kappa^2_BR}{\Lambda^4}\left(\lambda^{(0)}_\ell+\hf\right)\right]\nonumber.
\eea
These sums are formal because they are divergent. The regularization \eq{regul}
consists of including finite number of terms only, for which the arguments of the
logarithm functions are less then 1.

\section{Numerical results}
The main result of the computation above, the effective potential
\eq{sumsph}, will now be
used first to find out the value of the curvature or the homogeneous
conformal factor preferred energetically in the vacuum in the absence of 
bare cosmological constant, $\lambda_B=0$, and next
to obtain the induced, effective Newton and cosmological constants in
de Sitter spaces.

One usually introduces first counterterms which render the 
effective potential UV finite. Instead, we shall use the effective potential 
\eq{sumsph} without counterterms because our goal is first
to find the true vacuum at which the renormalization program,
the $R$-dependent counterterm set in particular, is supposed to be 
introduced after. Our procedure corresponds to the
minimization of the total action, the sum of the
finite pieces and the contributions of the counterterms.
We use the bare action in finding the saddle point of the path
integral which is always given in terms of the bare, cutoff theory.
Once the saddle point is found, then the divergences should
be removed in the standard fashion, by splitting the
bare action into the sum of renormalized and counterterm
parts. The minimization of the effective action will
be considered in this paper only. The renormalized
coupling constant and the counterterms will not be introduced,
instead the value of the curvature and the one-loop
improved effective coupling constants will be read of
at the saddle point.

\subsection{Effective potential}
It is easy to read off a {\em wrong} result from Eq. \eq{sumsph}.
When a gauge non-invariant cutoff is employed by cutting
off the summation over the spherical harmonics at a given
in the angular momentum value $\ell_{max}$ then on finds
\be
\gamma^{(1)ninv}(R)=C'\ln\frac{\kappa^2_BR}{\Lambda^4},
\ee
with $C'>0$. But one should use gauge invariant regularization which
requires in our case the definition of the cutoff in terms of the
eigenvalues of the Laplace operator. The difference between sharp cutoff
\eq{sharp} and the previous, non invariant regulator
is that as the curvature is increased the number of modes contributing to
the regulated path integral is decreased or remains constant, respectively.
This makes the effective potential changing faster with the curvature
when the gauge invariant regulator is used.

A qualitative estimate of the effective potential can be obtained
by exploiting the analogy with the Casimir effect for flat space,
summarized from the point of view of our problem in Appendix \ref{casimir}.
The curvature directly controls the volume of the de Sitter space
and indirectly, via the uncertainty principle, the level spacing
of the free particle energy spectrum. The former leads to the
Casimir effect when sharp cutoff is used and the latter
generates the same result for smooth cutoff.
The estimate of the one-loop effective potential is based on the
equipartition theorem which states that $\gamma(R)\approx{\cal N}\bar f$
where $\bar f$ is the typical contribution of a mode. The factor $\cal N$
is finite and well defined only if both IR and UV
sharp cutoffs are imposed and can easiest be obtained when the dimensions are
removed by the UV cutoff as in solid state physics or lattice field theory.
In our case ${\cal N}\approx\Lambda^4/R^2$.
In order to estimate $\bar f$ let us recall that the logarithm of
the Gaussian integral is the logarithm of the typical, average
value of the integral variable, apart of an uninteresting
$\ln2\pi$ term. Since the variables of the path integral
are identified by removing every dimension by the UV cutoff
the contribution of a mode is $\ln\kappa_B/\Lambda$ which gives
$\gamma(R)\approx c\Lambda^4/R^2\ln\kappa^2_B/\Lambda^2$.
Naturally this naive estimate, based on the typical contribution,
is reliable for ${\cal N}>>1$ only.

\begin{figure}
\includegraphics[height=14.cm,width=8.cm,angle=270]{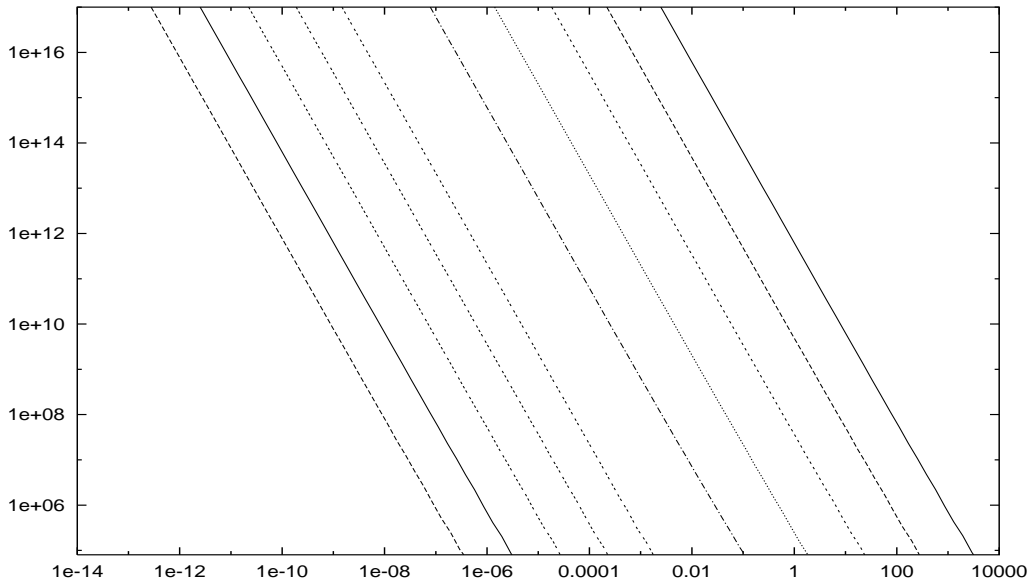}
\caption{The effective potential $|\gamma(R)|$ as a function
of $R/\kappa^2_B$, obtained with sharp cutoff. The lines belong
from left to right to $\Lambda^2/\kappa^2_B=10^{-5}$, $10^{-4}$,
$10^{-3}$, $10^{-2}$, $10^{-1}$, $1$, $10$, $10^2$, $10^3$, $10^4$
and $10^5$. Since $\gamma(R)<0$ in the given range of curvature
for $\Lambda^2/\kappa^2_B\ge1$ it is $-\gamma(R)$ which is shown for
these values of the cutoff. The slightly bigger distance between
the curves of $\Lambda^2/\kappa^2_B=0.1$ and
$\Lambda^2/\kappa^2_B=1$ when the
sign changes witnesses the logarithmic function in Eq. \eq{copi}.
\label{rf}}
\end{figure}

\begin{figure}
\includegraphics[height=14.cm,width=8.cm,angle=270]{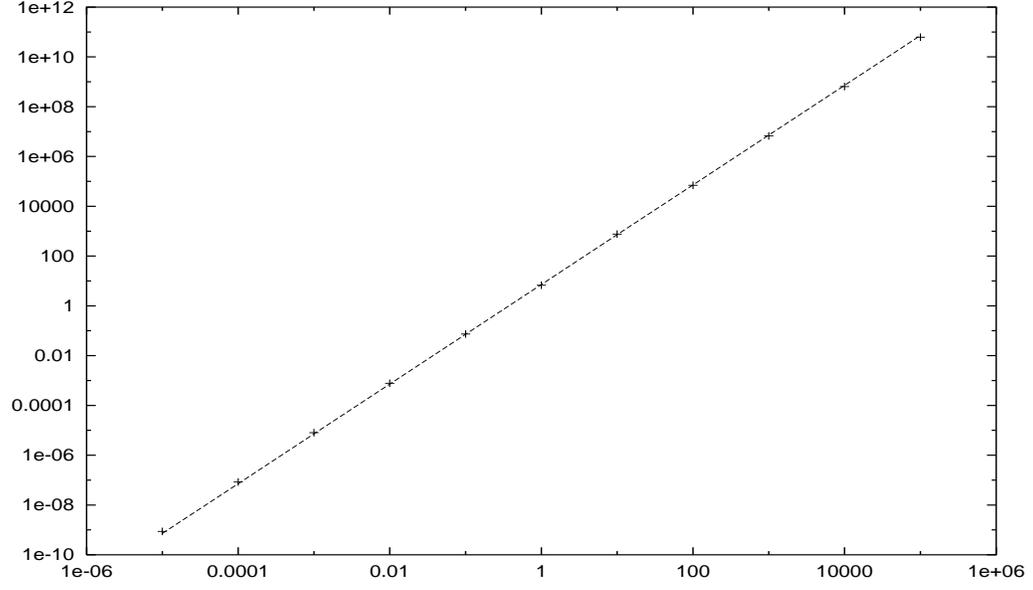}
\caption{The numerical and the fitted values for
$M^4(\Lambda)/\ln c_2\kappa^2_B/\Lambda^2$ are shown
by dots and continuous line, respectively, as functions
of $\Lambda^2/\kappa^2_B$.\label{plm}}
\end{figure}

The numerical evaluation of the the effective potential obtained by means
of the sharp cutoff is depicted in Fig. \ref{rf}. The highest value
of $\ell$ contributing to the effective potential
\eq{sumsph} is approximately $4\cdot10^4$ (small curvature) and
$10$ (high curvature) along the curves of this Figure.
The flat space Casimir effect for a dimensionless scalar field
suggests the form
\be\label{ueff}
\gamma^{(1)}(R)=\begin{cases}M^4(\Lambda)
\left(\frac1{R^2}-\frac1{R^2(\Lambda)}\right)&R\le R(\Lambda)\cr
0&R>R(\Lambda)\end{cases}
\ee
with
\be\label{mnegy}
M^4(\Lambda)=c_1\Lambda^4\ln\frac{c_2\kappa^2_B}{\Lambda^2},
\ee
and $R(\Lambda)=c_3\Lambda^2$. The second term in the parentheses on the
right hand side of Eq. \eq{ueff} is to cancel the effective potential
for high enough curvatures when no eigenvalues are left in the sum by the 
sharp cutoff. The effective potential is naturally not a smooth function for
$R\approx R(\Lambda)$ anymore and one should not consider the
theory in this regime where few modes are left only. The fit of
the numerical results yields $c_1\approx7.201$, $c_2\approx2.989$ and
$c_3\approx0.665$, c.f. Fig. \ref{plm}.

Let us now consider the complete effective potential
\be\label{copi}
\gamma(R)=-\frac{v\kappa^2_B}{R}+c_1\Lambda^4
\left(\frac1{R^2}-\frac1{R^2(\Lambda)}\right)
\ln\frac{c_2\kappa^2_B}{\Lambda^2},
\ee
where
\be\label{terfogat}
V=\int dx\sqrt{g}=\frac{v}{{R}^2}
\ee
is the four volume of the de Sitter space with $v=3200\pi^2/3$.
One finds a quantum phase transition
at $\kappa^2_B=\kappa_{cr}^2=\Lambda^2/c_2$
when $R_{min}$, the curvature where $\gamma(R)$ reaches its minimum
changes in a discontinuous manner. Deeply inside of the small cutoff
phase, $\kappa_{cr}^2<<\kappa^2_B$, the energetically favored curvature is
\be
R_{min}=\frac{2c_1\Lambda^4}{v\kappa^2_B}\ln\frac{c_2\kappa^2_B}{\Lambda^2}
\ee
For large cutoff, $\kappa^2_B<\kappa_{cr}^2$, the minimum of the
effective potential is reached at $R_{min}=0$.

It is easy to reconstruct the effective potential for the homogeneous
conformal mode, $\phi(x)=\Phi$, too. In fact, all needed
according to Eq. \eq{conff} is the replacement $R\to R/(1+\Phi)$
in the effective potential. Therefore the small or the large cutoff
phase prefers $\Phi=R/R_{min}-1$ or $\Phi=\infty$, respectively.
But on should keep in mind that we can not settle the problem
of the instability by studying the dynamics of the homogeneous
conformal mode only. In fact, the instability of the
inhomogeneous and the homogeneous conformal modes comes
from the kinetic energy (first) or the potential energy (second)
term on the right hand side of Eq. \eq{actcon}. All one can say
is that as long as the theory is IR finite the homogeneous
mode has a useful diagnostic role, the long wavelength
inhomogeneous modes follow a dynamics which is similar to those of
the homogeneous mode.

\subsection{Induced cosmological constant}
The effective potential \eq{copi} suggests the effective action
\be\label{ansatzc}
\Gamma[g]=-\kappa^2_{eff}\int dx\sqrt{g(x)}F(R(x)),
\ee
where $F(R)$ is a polynomial of the curvature. In order to define
a unique coupling constant $\kappa^2_{eff}$ we impose the condition
\be
\frac{dF(R)}{dR}_{|R=0}=1
\ee
on $F(R)$. The inspection of Eq. \eq{copi} yields
$F(R)=R-2\lambda-gR^2$ with $\kappa^2_{eff}=\kappa^2_B$,
\be\label{kozmk}
\lambda=\frac{c_1\Lambda^4}{2v\kappa^2_B}\ln\frac{c_2\kappa^2_B}{\Lambda^2}
\ee
and
\be
g=-\frac{c_1\Lambda^2}{2c_3v\kappa^2_B}
\ln\frac{c_2\kappa^2_B}{\Lambda^2}
\ee
for $R<<\Lambda^2$. Notice that the fourth power of the cutoff
in Eq. \eq{kozmk} leads to a constant counterterm by dimensional
reason. It indicates that the vacuum fluctuations contribute
to the action in a homogeneous manner in space-time which is a
manifestation of the Casimir effect, known from the normal ordering
prescription in Quantum Field Theory. The one-loop quantum fluctuations
leave the Newton constant unchanged but generate a cosmological term and
a higher derivative piece.

We see that the difference between the small and large cutoff phase is
that the induced cosmological constant is positive and negative,
respectively. Furthermore there is a higher order derivative term
generated with coupling of the opposite sign as the cosmological constant.
The expressions \eq{quadr} and \eq{sumsph} indicate
that the de Sitter background is unstable in the one-loop effective
theory for $\lambda>cR$ where
\be
c=\min\left(\frac{\lambda_\ell^{(2)}}{2}+\frac13,
\frac{\lambda_\ell^{(1)}}{2}+\frac18,\frac{\lambda_\ell^{(0)}}{2}\right)
=\hf.
\ee
The low cutoff phase is therefore unstable for $R<R_{min}/2$
and the entire large cutoff phase is stable as expected.

\subsection{Running Newton constant}\label{rnc}
In order to shed more light onto the nature of the two phases
we introduce the effective strength of interaction by using the ansatz
\be\label{ansatz}
\Gamma[g]=-\int dx\kappa^2(R(x))\sqrt{g(x)}R(x),
\ee
where the space-time dependence is explicitly shown. The effective
Newton constant, found by comparing Eqs. \eq{copi} and \eq{ansatz},
\be\label{efnc}
G(R)=\frac1{16\pi\kappa^2(R)}
=\frac1{16\pi\kappa^2_B}\frac1{1-\frac{c_1\Lambda^4}{v\kappa^2_B}
\left(\frac1{R}-\frac{R}{R^2(\Lambda)}\right)\ln\frac{c_2\kappa^2_B}{\Lambda^2}}
\ee
is a measure of the strength of interactions. In fact,
$\kappa^2(R)/\kappa_B^2$ measures the suppression of small fluctuations
of the metric tensor, $\Delta g$, compared to the classical Einstein theory with
vanishing cosmological constant. The fluctuations of the one-loop
effective theory are suppressed at length scales $\ell>>\sqrt{G}$ where
$\sqrt{G}$ "seems small". In fact, simple order of magnitude estimate gives
$\Delta g\approx\ell_{Pl}/\ell$ for the typical fluctuations
of the metric tensor of length $\ell$ where $\ell_{Pl}=\sqrt{G}$
denotes the renormalised Planck length. The running Newton constant
has a central importance in the phenomenology of quantum effects in gravity.
A distance dependent Newton constant may provide a solution to the astrophysical 
missing mass problem \cite{rute,boue}.

The tree level theory is weakly coupled when the cutoff is increased as long as we 
stay in the small cutoff phase. Only the modes beyond the Planck energy
which appear in the large cutoff phase are strongly coupled.
This feature is changed drastically by the leading order loop corrections
because the one-loop theory on the de Sitter space of curvature $R$
is perturbative for length scales
\be
\ell^2>\ell^2_{Pl}(R)=\frac{\ell^2_B}
{1+\frac{16\pi c_1}{v}\Lambda^4\ell^2_B
\left(\frac1{R}-\frac{R}{R^2(\Lambda)}\right)
\ln\frac{16\pi\ell^2_B\Lambda^2}{c_2}},
\ee
where $\ell^2_B=1/16\pi\kappa^2_B$ 
and the whole theory is perturbative if $\ell_{Pl}\Lambda<1$,
\be
\Lambda^2\ell^2_B<1+\frac{16\pi c_1}{v}\Lambda^4\ell^2_B
\left(\frac1{R}-\frac{R}{R^2(\Lambda)}\right)
\ln\frac{16\pi\ell^2_B\Lambda^2}{c_2}.
\ee

In order to simplify the discussion we shall restrict ourselves
into regions far away form the transition region, deeply in the small and
large cutoff phases, for $\Lambda^2<<\ell^{-2}_B$ and
$\ell^{-2}_B<<\Lambda^2$, respectively.
There is an IR Landau pole in the small cutoff phase at $R_L=R_{min}/2$
and $G(R)$ is decreasing from infinity to $G$
as $R$ is increased from $R_L$ to $R(\Lambda)$. At the energetically
preferred curvature we find $G(R_{min})=2G$ and
$\ell^2_{Pl}(R_{min})=2\ell^2_B$. All modes are non perturbative
and the theory strongly coupled unless $R>R_L$. The modes become
perturbative rapidly as $R$ is increased beyond the Landau pole and
the entire theory is perturbative for $R_{min}\le R$.
The running Newton constant is an increasing function of the curvature
in the large cutoff phase and one finds
\be\label{nlir}
G(R)\approx\frac{Rv}{16\pi c_1\Lambda^4\ln\frac{16\pi\ell^2_B\Lambda^2}{c_2}}
\ee
for $R<R(\Lambda)$ and $G(R(\Lambda))=G_B$. In a similar manner
$\ell_{Pl}(R)$ increases with $R$, $\ell^2_{Pl}(R)\approx R/\Lambda^4$
as long as $R<R(\Lambda)$ and $\ell^2_{Pl}(R(\Lambda))=\ell^2_B$.
All modes appear perturbative in the effective theory for $R<<\Lambda^2$.

The peculiar sensitivity of quantum gravity on the UV sector is
reflected in the expression \eq{efnc}, too.
When the running coupling constant is computed in flat space-time in
the presence of a background field, such as in Yang-Mills theories
\cite{savvidy}, the cutoff appears in the argument of the logarithmic
function only. As mentioned in connection of Eq. \eq{kozmk},
that the factor $\Lambda^4$ in front of the logarithimc function
implies a constant, field independent counterterm, the hallmark of the
Casimir effect.

\section{Dynamics of the conformal modes}
The unstable conformal modes have been kept frozen at $\Omega=1$ in the 
one-loop computation. The complete functional integral can be obtained
by factorizing the integral measure for 
$g_{\mu\nu}=g^{(R)}_{\mu\nu}(1+\phi)+h'_{\mu\nu}$, where 
$h'$ is obtained by retaining the first three contributions only in the 
right hand side of Eq. \eq{reduct}, as $D[g]=D[\phi]D[h']$. The regulator
is defined as before by means of the eigenvalues of the $\Box$ operator.
The integration over $h'$ in Eq. \eq{bfpi} yields
\be
e^{W[\phi,j]}=\int D[h']e^{-S[g]+\int dxj^{\mu\nu}g_{\mu\nu}}.
\ee
The Legendre transformation for a given $\phi$ configuration is defined by
\be\label{filetr}
\Gamma[\phi,g]=-W[\phi,j]+\int dxj^{\mu\nu}g_{\mu\nu}
\ee
with
\be
g_{\mu\nu}=g^{(R)}_{\mu\nu}(1+\phi)+h'_{\mu\nu}=\frac{\delta W[\phi,j]}{\delta j_{\mu\nu}},
\ee
cf. Eqs. \eq{effaa} and \eq{geffaa}. 

Notice that the dependence on $\phi$ enters explicitly and through $g$ in the
Legendre transform \eq{filetr}. The effective actions appearing in Eqs.
\eq{ansatzc} or \eq{ansatz} are actually $\Gamma[\phi,g]$, computed at $\phi=0$.
The essence of the anstaz is the upgrading of a simple one-loop result 
obtained on a fixed de Sitter geometry to a gauge (diffeomorphism) invariant 
functional. We extend this procedure by requiring the usual transformation
rules under conformal transformations, too. This assumtion suppresses the explicit 
dependence on $\phi$ and gives rise to
\be\label{teljesw}
e^{W[j]}=\int D[\phi]e^{-S_c[\phi]+\int dxj^{\mu\nu}g_{\mu\nu}}
\ee
with $S_c[\phi]=\Gamma[0,g]$ for the complete generating functional.

The effective action $S_c[\phi]$ is simpler in terms of 
$\Omega^2=1+\phi$ and the functional, taken from Eq. \eq{ansatzc}, leads to
\be
S_c[\phi]=\kappa^2\int dx\sqrt{g}\left[6(1-2gR\Omega^{-2})\Omega\Box\Omega
+36g\Omega^{-2}(\Box\Omega)^2+V(\Omega)\right]
\ee
according to Eq. \eq{resc}, where
\be
V(\Omega)=-R\Omega^2+2\lambda\Omega^4+gR^2.
\ee
The ansatz \eq{ansatzc} and the assumption about the proper
conformal properties amount to a partial resummation of the loop-expansion
for the stable modes and the dependence of the effective action on the conformal
factor $\Omega$ is supposed to become more reliable.

It is easy to identify the stabilization mechanism (ii) mentioned in the
Introduction in the low cutoff phase where the potential $V(\Omega)$ is bounded 
from below ($\lambda>0$). We write $\Omega=\bar\Omega+\omega$, where the potential fixes 
$\bar\Omega^2=R_{min}/4\lambda=1$. The instability of the inhomogeneous modes 
is still visible because both the $\ord{\Box}$ and  the $\ord{\Box^2}$ terms are 
negative definite because $g<0$. But the increase of
the wave number is not a real instability in theories with finite cutoff.
In fact, models with unstable dispersion relation are stabilised
by the UV regulator and give rise stable, modulated vacuum \cite{afvac} whose
characteristic scale is of the cutoff. The dispersion relations for the
excitations above such a vacuum have several branches as in solids. The
heavy modes, optical phonons, decouple from the low energy dynamics
which is dominated by the light branches, acoustic phonons,
and the inhomogeneity of the vacuum is not resolved by measurements
with finite resolution for sufficient large values of the cutoff.

But the fate of the high cutoff phase remains unclear. The potential 
$V(\Omega)$ in unbounded from below and the increase of the amplitude of 
$\Omega$, the decrease of the curvature and the increase of the volume,
remains as instability of the one-loop dynamics. The complication here
is that the prefered vacuum of the one-loop dynamics, $R=0$, can not be 
reached in the scheme presented here which is valid for $R>0$. Furthermore,
fluctuations around $R=0$ explore the negative curvature geometries, as well,
where the dynamics is not given by the analytical continuation of our formulae.

\section{Summary and conclusion}
We argue in this paper that the instability of Euclidean
Einstein gravity is an indication of a non trivial vacuum
structure whose description requires non perturbative
methods. As a simple step in this direction the one-loop renormalization
was considered on an Euclidean de Sitter background geometry by
suppressing the unstable modes. The similarity of this problem with the
Casimir effect was emphasized together with the role of the curvature
as an IR cutoff which makes quantum gravity specially sensitive to the
UV regime. The effective action was minimized within the family of
homogeneous spaces in order to identify the vacuum.

Two phases were found and the effective strength of interaction
behaves in a manner which is contrary to what the tree-level
dynamics would suggest. In the small cutoff phase,
$\Lambda<<1/\sqrt{\ell_B}$, the conformal modes are stabilised by the one-loop
effective action for the other tree-level stable modes and we find a close analogy 
with the Savvidy vacuum of Yang-Mills theories. The system prefers a 
non vanishing curvature, $R_{min}\approx\Lambda^4\ell^2_B$ and $\ell_{Pl}=\sqrt{2}\ell_B$.
The tree-level theory is weakly coupled for arbitrary curvature
but the fluctuations acquire large amplitude at length
scales $\ell>1/\sqrt{R_{min}}$ and the theory is weakly coupled for
$R_{min}\le R$ only when these modes are excluded due to the smallness
of the space-time volume. Deeply in the large cutoff phase when the
cutoff is increased far beyond the transition point then the minimum of
the effective action is reached at $R_{min}\approx0$, the fluctuations 
are suppressed. But the stability of the conformal modes remain an open
question.

We proposed two parametrizations for the effective action.
The running Newton constant can be defined in the effective theory
without cosmological term which amounts to assuming formally massless
quantum fluctuations. This coupling constant determines the length scale
below which the quantum fluctuations are strong. Another
possibility is to allow a radiative correction induced cosmological term
explicitly in the effective action. This choice is motivated by the
proportionality of the Casimir effect with the space-time volume and is
more physical because it produces slower dependence on $R$ in the
effective parameters than the previous case. Furthermore, a higher
order derivative term is recovered by means of this parametrization.
The effective Newton constant agrees with the bare one in this scheme
and the two phases are distinguished by the sign of the induced
cosmological constant and the coefficient of the higher order derivative term.

The change of the UV cutoff while the other parameters are kept
constant modifies the physical content of the theory and
even drives the system through a phase transition. The cutoff dependence
in general is not surprising in a nonrenormalizable theory. It might
turn out that some so far unknown nonperturbative effects render the
Einstein-Hilbert theory renormalizable but in lacking these elements
one has to accept that the theory which is nonrenormalizable in
the framework of the loop-expansion produces cutoff-dependent results
at one-loop level. Our conclusions drawn correspond to energy scales
well below the cutoff and are therefore independent of the details of
the regularization procedure since this latter is always represented
by irrelevant operators.

The cutoff dependence of nonrenormalizable
models usually remains quantitative. The strong cutoff-effects reported
in this work can be traced back to the different sign of the
induced cosmological constant in the two phases. This parameter is
renormalizable and its sign influences the physics in a more substantial,
quantitative level.

We encountered two level of instabilities in this work. In the tree-level
theory the conformal modes are unstable but they seem to be
stabilized by the quantum fluctuations at least in the long distance
region of the high cutoff phase. The one-loop improved effective
theory remains unstable in the low cutoff phase. The cosmological
constant appears formally as the parameter $-m^2$
in a scalar model. The instability of the low cutoff phase
with  $\lambda>0$ should therefore be similar to those of a
scalar model with spontaneously broken symmetry. In the latter case
the tree-level contributions to the renormalization group
equation were found to be crucial \cite{trrg} in order to
establish a systematical approach to the mixed phase of
first order phase transitions, in particular the spinodal phase
separation \cite{spinod} where the soft modes arising from
the breakdown of space-time symmetries by the inhomogeneous
saddle points generate strong nonperturbative effects, among others
the Maxwell-cut.

We continue with few comments on the results of the previous Section
from the point of view of the case (ii) of the stabilization mechanism,
mentioned in the Introduction and identified in the low cutoff phase. 
It was found that the one-loop effective action
may stabilize the vacuum and generate a condensate of the metric tensor
even in the absence of matter. It remains to be seen if such a condensate
involves the conformal modes only or the spin 2 components, too.
Another open question is whether the new vacuum with condensate is weakly or
strongly coupled. The dimensionless ratio $\kappa_B/\Lambda$
available in this nonrenormalizable theory may generate
small or large dimensionless numbers and this question can be settled
by detailed computation only. Though the other stabilization mechanisms
were not pursued in this work it is clear that the excitations above
the condensed vacuum produced by either mechanism listed in
the Introduction are different than those above
the naive, 'empty' vacuum and the success of general
relativity in the classical domain requires more careful
justification.

One arrives at a surprising conjecture by following the analogy
offered by the scalar model. It is well known that the speed of sound
is vanishing in the mixed phase. This is due to the strongly
coupled soft modes, the translations of the domain walls. In more
physical terms the domain walls absorb the sound. In a similar manner
one expects the speed of the propagation of gravitons be vanishing
in the low cutoff phase. This situation where the naive plane waves
ceases to propagate and drop out from the asymptotical sector
is called confinement. If this mechanism is realized then
gravitons will be confined in the low cutoff phase due to the quantum
liquid nature of the vacuum. The confinement radius
which can be approximated by the length scale of the IR Landau pole
is in the order of magnitude of the size of the stable de Sitter space,
indicating simply that the gravitons should be confined by the horizon.

Finally we mention other natural extensions of this work, the inclusion of 
matter fields and the cosmological constant in the bare action. 
Condensate modifies the vacuum structure and it may stabilise the 
gravitational sector by breaking conformal invariance. The interplay of condensates
in the coupled Einstein-Yang-Mills system is a particularly interesting problem. 
The bare cosmological constant, included from the very beginning in bare action 
should preserve the topology of the phase structure but make the result
more realistic. We plan to report on such results soon.

\begin{acknowledgments}
We wish to acknowledge a stimulating discussion with G. Veneziano.
\end{acknowledgments}

\appendix
\section{Casimir effect in a box}\label{casimir}
It is instructive to consider a dimensionless, real, free, massless
scalar field $\phi$ in a box of linear size $L$ of the
Euclidean space-time subject of periodic boundary conditions.
The generator functional for connected Green functions is
\be\label{szabadw}
e^{W[j]}=\int D[\phi]e^{\frac{\mu^2}{2}\int dx\phi_x\Box\phi_x
+\int dxj_x\phi_x}.
\ee
The effective action
\be
\Gamma[\phi]=-\frac{\mu^2}{2}\int dx\phi_x\Box\phi_x+\hf\tr\ln\mu^2\Box
\ee
takes the value
\be
\Gamma[0]=\hf\tr\ln\mu^2\Box
\ee
in the vacuum.

The evaluation of the formal functional trace requires regularization.
We choose lattice regularization and write
\be\label{fourier}
\phi=\frac{a^2}{L^2}\sum_nc_ne^{i\frac{2\pi}{L}n_\mu x^\mu},
\ee
where $a$ denotes the lattice spacing and the summation is restricted for
\be
0<n^2\frac{(2\pi)^2}{L^2}<\Lambda^2=\left(\frac{2\pi}{a}\right)^2
\ee
and the regulated integral measure is given by
\be\label{dimtlan}
D[c]=\prod_ndc_n.
\ee
One finds finally
\be\label{partfa}
\Gamma[0]=\hf\sum_{n^\mu\not=0}\Theta\left(\Lambda^2
-n^2\frac{(2\pi)^2}{L^2}\right)\ln\left(\mu^2n^2
\frac{(2\pi)^6}{L^2\Lambda^4}\right).
\ee
One can simplify this expression for $1<<\Lambda L$ as
\bea\label{contlfr}
\Gamma[0]&=&L^4\int_{|p|\le\Lambda}\frac{d^4p}{(2\pi)^4}
\ln\frac{p^2\mu^2(2\pi)^4}{\Lambda^4}\nn
&=&\frac{L^4\Lambda^4}{32\pi^2}\left(\ln\frac{\mu^2(2\pi)^4}{\Lambda^2}-\hf\right).
\eea

\end{document}